\documentclass[twoside]{article}
\usepackage[preprint]{paperstyle}
\makeatletter
\renewcommand{\Notice@String}{$^*$ Equal contribution.}
\gdef\@runningauthor{Boris Velasevic, Eden Saig, Eric Mazumdar}
\def\ftype@copyrightbox{8}
\def\@copyrightspace{
\@float{copyrightbox}[b]
\begin{center}
\setlength{\unitlength}{1pc}
\begin{picture}(20,0.5)
\put(0,1.5){\line(1,0){4.818}}
\put(0,0){\parbox[b]{19.75pc}{\small \Notice@String}}
\end{picture}
\end{center}
\end@float}
\makeatother

\usepackage{bbm, algorithm, algorithmic,hyperref,amsmath}
\usepackage{url}
\usepackage{graphicx}
\usepackage{enumitem}

\usepackage{amsmath,amsfonts,bm}

\def\eqref#1{equation~\ref{#1}}

\def\1{\bm{1}}

\def\vg{{\bm{g}}}

\def\vs{{\bm{s}}}

\DeclareMathAlphabet{\mathsfit}{\encodingdefault}{\sfdefault}{m}{sl}
\SetMathAlphabet{\mathsfit}{bold}{\encodingdefault}{\sfdefault}{bx}{n}

\def\gB{{\mathcal{B}}}

\def\gG{{\mathcal{G}}}
\def\gH{{\mathcal{H}}}

\def\gP{{\mathcal{P}}}
\def\gQ{{\mathcal{Q}}}

\def\gS{{\mathcal{S}}}
\def\gT{{\mathcal{T}}}

\newcommand{\R}{\mathbb{R}}

\DeclareMathOperator*{\argmin}{arg\,min}

\usepackage[dvipsnames]{xcolor}
\usepackage{xurl}
\usepackage{cleveref}

\hypersetup{
    colorlinks,
    linkcolor={red!50!black},
    citecolor={blue!50!black},
    urlcolor={blue!80!black}
}

\newtheorem{theorem}{Theorem}

\newtheorem{proposition}{Proposition}

\newcommand{\Set}[1]{\left\{{#1}\right\}}

\newcommand{\indicator}[1]{\mathbbm 1 \Set{{#1}}}
\newcommand{\tv}[2]{\operatorname{d_{TV}}\left({#1},{#2}\right)}

\renewcommand{\eqref}[1]{Eq.~(\ref{#1})}

\usepackage[numbers]{natbib}

\begin{document}

\twocolumn[

\papertitle{
Anytime Contracts for AI Services
}

\paperauthor{
Boris Velasevic$^*$
\And
Eden Saig$^*$
\And
Eric Mazumdar}

\paperaddress{ \\California Institute of Technology } ]

\begin{abstract}
    
AI services are prone to moral hazard, as model providers can silently route requests to cheaper, lower-quality models, and the inherent stochasticity of model outputs makes this degradation hard to detect. Contract theory addresses moral hazard by designing payment rules that reward high-quality outputs and prevent strategic overcharging. However, existing frameworks assume interactions of fixed duration, whereas AI services are typically consumed through subscriptions and pay-as-you-go APIs that users can terminate at any time based on observed performance. We show that this flexibility can break standard contracts, opening them up to abuse by either the providers or users. To address this, we extend contract theory to account for unknown exit rules. We formalize the problem as a Principal-Agent game, and show that budget-optimal contracts in this setting are proportional to anytime-valid hypothesis tests based on likelihood-ratio e-values. We give an efficient algorithm for computing anytime contracts and show how betting e-processes can be used to approximate them when the models are unknown. Finally, we evaluate our findings on synthetic data and data from deployed LLMs.

\end{abstract}

\section{Introduction}

 AI systems are becoming ubiquitous and increasingly embedded in high-stakes decision-making across domains. Such decisions often require the intricate reasoning abilities of high-quality models, which most organizations cannot train and operate in-house, leading them to delegate AI tasks to external service providers. Due to their generality, AI systems are increasingly viewed as \emph{tools}, used for indefinite durations within organizational processes.
 Accordingly, payment for such services typically takes the form of auto-renewing contracts or pay-as-you-go APIs, which are arrangements without a predetermined end.

Together with growing demand, the increasing costs of high-quality models create economic risks.
Both model inference and the evaluation of generative-model outputs are inherently noisy, while the process by which providers route queries to different models is typically opaque.
This creates \emph{information asymmetry} between AI service providers and their users wherein an end-user cannot tell whether a low quality outcome is the result of randomness or deliberate degradation from the service provider. Since higher-quality models are generally more costly to operate, profit-maximizing providers can benefit from routing requests to cheaper, lower-quality models. This gives rise to  \emph{moral hazard}~\citep{saig2024incentivizing,cai2025,velasco2025your}.
We note that incidents of quality degradation are well-documented in practice \citep{chen2023chatgpt, gao2025model}, and it's not hard to believe that fully strategic providers will leverage such gaps to their advantage.

Moral hazard is a classic problem in economics, commonly addressed by designing pay-for-performance payment rules, encoded as \emph{contracts}, which reward the agent for quality signals that indicate effort \citep{holmstrom1979, grossmanhart1983, dutting2024algorithmic}. Recent work has started applying these ideas in settings native to machine learning \citep{saig2024incentivizing,saig2026adaptive}, designing payment schemes that implicitly assume fixed-duration interactions.
However, when AI services are used on an ongoing basis, contractual interactions become inherently \emph{dynamic}.
Payment schemes used in practice either have an indefinite duration by design (e.g., APIs) or contain an exit clause that can be triggered at any time by either party, potentially mandated by regulators \citep{eu2011consumer}.
For example, one may want to cancel a subscription or switch to a different provider.
In such settings, routing decisions interact with concerns about contract duration, an interplay which may cause fixed-duration contracts to achieve an opposite goal from the one they target.

To illustrate this, consider a setting in which, at each step, a strategic provider receives a task from a client organization and routes it to either a costly high-quality model or a cheaper low-quality one, then each task either succeeds or fails.
In the fixed-duration setting, the optimal contract incentivizing high effort for the minimal expected pay is the one that pays a large amount only if all tasks are successful (e.g., \citep{dutting2019simple, bond2009multitask}). Such contracts fail to incentivize high effort in our setting: if the client terminates after the first failure, payment is always zero at termination, so a provider anticipating such behavior has no incentive to bear the cost of the higher-quality model and therefore only offers lower-quality outputs.

Alternatively, another common approach is to pay per successful task. However, if the client only cares about completing a task and terminates once it succeeds---for example, when trying to prove a major mathematical theorem---the provider receives exactly one payment regardless of duration. Hence, when a substantially cheaper model has some chance of success, the provider may therefore prefer to route prompts to it instead.

Both examples show that exit clauses can create economic failures under different stopping criteria. Since the rules governing whether interaction continues or terminates are varied and typically unknown in advance, there is a need for tools that explicitly account for these dynamics and their uncertainty.

\paragraph{Contributions:} In this work, we develop a theory of contract design for AI services that explicitly accounts for the dynamic nature of the problem and---crucially---the exit clauses that are key features of AI services. Our contributions are as follows.

\begin{itemize}[leftmargin=*]
    \item \textbf{Contracts under stopping time uncertainty:} We formalize the interaction as a Principal-Agent game in which the provider deploys a potentially adaptive model-routing policy, quality is measured at each step, and the user may choose to terminate the contract according to any---possibly data-dependent---stopping rule in a prescribed class $\mathcal{T}$. We show that accounting for exit clauses can cause existing contract-theory approaches to fail, motivating the need for contracts that explicitly deal with stopping time uncertainty.
    In the resulting formulation,  we require that the provider prefers providing the high-quality policy over their alternative uniformly over all stopping rules a user could choose. This reflects the fact that a user's stopping rule is typically private and could be arbitrary. The resulting formulation is a strict generalization of existing fixed-duration contract design problems in machine learning which implicitly consider the case where the class of admissible stopping rules is a singleton.
    \item \textbf{Anytime contracts:} Our central result characterizes minimum-budget contracts robust to stopping-time uncertainty.
   The optimal contract takes the form of a scaled sequential likelihood-ratio test on the sequence of observed quality signals, with the required budget increasing when the target policy is more costly or harder to distinguish statistically. The resulting likelihood-ratio processes are also log-optimal e-processes \citep{grunwald2024safe,ramdas2025hypothesis}, extending the connection between statistical testing and contract design \citep{saig2023delegated,saig2024incentivizing} to anytime-valid inference \citep{shafer2021testing,ramdas2023savi}. For bounded stopping time families, we also show that these contracts can be computed efficiently by reducing
    to an optimal stopping problem.
    \item \textbf{Audits with partial information.}
    Implementing the optimal anytime contract requires
    knowledge of the output distributions induced by the provider's routing policies. Using a standard e-value
    construction \citep{waudbysmith2024}, we develop an approximation to the optimal contract that can be implemented
     from observed signals alone, provides anytime
    service-level guarantees, and can sometimes recover the optimal anytime contract exactly.
    \item \textbf{Empirical validation under realistic conditions:} We present empirical validation of our theory both in a simple synthetic environment and with real data from closed-source LLMs. Synthetic experiments illustrate how stopping-time uncertainty changes optimal payments and worst-case stopping behavior. Experiments using the Holistic Agent Leaderboard \citep{kapoor2026hal} examine robustness costs, partial distributional information, and demonstrate how contracts can deter multiple alternative policies.
\end{itemize}

\subsection{Related Work}\label{sec:related}
\paragraph{Moral hazard in {AI} delegation.}
Our work connects to an emerging line of research on moral hazard in delegated AI settings. Notably,
\citet{saig2024incentivizing} develop a contract design framework for delegated text generation in fixed-duration settings with cost uncertainty;
\citet{velasco2025your} study incentives under token-based billing;
\citet{cai2025} audit whether an API in fact serves the model it
advertises;
and \citet{saig2026adaptive} design adaptive contracts that account for costly evaluation. We extend this line of work by identifying and addressing failure modes arising due to the open-ended nature of AI use, applicable to settings such as monthly subscriptions and pay-as-you-go APIs.

\paragraph{Statistical contracts.}
\citet{saig2023delegated, saig2024incentivizing} establish a
correspondence between optimal contracts and optimal statistical tests for
delegated learning and text generation. Their formulations assume a fixed duration, while we study how the design problem changes
when the contract is allowed to terminate at an
observation-dependent stopping time. \citet{bates2026principal} connect
e-values to optimal menus in statistical testing settings with adverse selection, where the provider's type is uncertain. In contrast, we consider general \emph{moral hazard} settings, in which the provider's type is known but their action is hidden, and show that likelihood-ratio e-processes arise as optimal contracts.

\paragraph{Dynamic contracting.}
Within the broader moral hazard literature, our work contributes to the study of dynamic contracting, where contractual relationships have a temporal component possibly evolving over time (e.g., \citep{holmstrom1987aggregation,fudenberg1990short,cvitanic2012contract}).
Our formulation is close in spirit to that of \citet{ezra2026contract}, who study contracts with sequential actions and stopping criteria strategically set by the agent. In contrast, our setting assumes that the stopping rule is chosen adversarially, and incentive compatibility must hold uniformly over an admissible class of stopping times.

\section{Problem Setting}\label{sec:setting}

\newcommand{\llm}{{g}}
\newcommand{\llms}{{\gG}}
\newcommand{\router}{{\pi}}
\newcommand{\contract}{{w}}
\newcommand{\prompt}{{x}}
\newcommand{\response}{{y}}
\newcommand{\signal}{{s}}
\newcommand{\evaluator}{{q}}

\begin{figure*}[!t]
    \includegraphics[width=\textwidth]{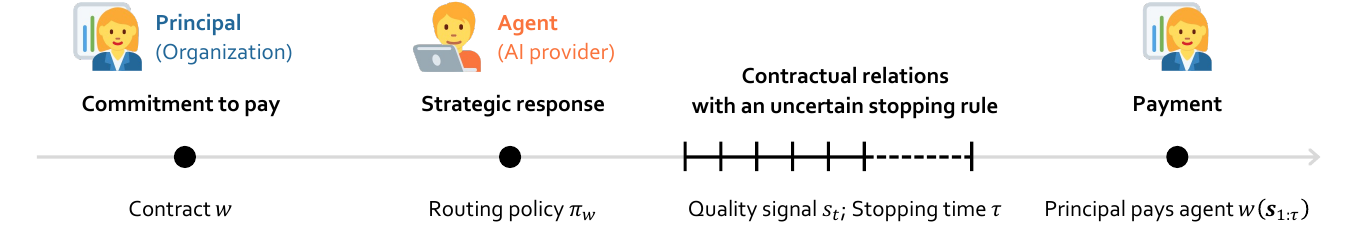}
    \caption{
    Contracts with stopping rule uncertainty as a Principal-Agent game (\Cref{sec:setting}).
    }
    \label{fig:schematic}
\end{figure*}

We formalize our setting as a Principal-Agent game, extending the classic contract design framework to accommodate \emph{stopping-time uncertainty}.
Our setting is illustrated in \Cref{fig:schematic}.

The game consists of two players: a \emph{client organization} (Principal) who seeks
access to AI services and has bargaining power to offer performance-based
pay, and a \emph{provider} (Agent) who operates
a set of generative
models at varying levels of quality and cost.
Interaction takes place over discrete steps $t \in \{1,2,\dots\}$: at
each step, the client sends one or more prompts $x \sim D$ and receives a
response $y$ for each. Importantly, we make no assumption on the amount or frequency of data collected at each step. A step may correspond to a single query, a batch of interactions aggregated over a month, or any other, potentially irregular, unit of observation. Interaction may terminate after each step based on the observed quality of outputs up to that step. Our goal is to design an \emph{anytime} payment scheme which incentivizes the provider to invest effort despite this uncertainty.

To generate responses, the provider operates a set of $k$ generative models, denoted by $\gG$. Each model $g\in\gG$ is a mapping from a prompt $x\sim D$ to a response $y=g(x)$.
We assume that each call to a model $g$ has expected cost $c_g\ge0$.
For each prompt, the model generating the response is picked by a
\emph{router} $\pi$, which maps prompts to generators,
possibly at random and adaptively based on past performance\footnote{Formally, the routing decision $\pi(t)$ is measurable with respect to $\sigma(x_1, y_1, ..., x_{t - 1}, y_{t - 1}, x_t)$ augmented by additional private randomization.}.
The client organization seeks generated responses from a high-quality routing policy $\pi_\mathrm H$, rather than a low-quality routing policy $\pi_{\mathrm L}$. We study the two-action setting --- a provider makes a decision which of the two routers to deploy. Empirically, we observe that our contracts can sometimes deter a much larger class of actions.

\paragraph{Moral hazard in AI services.} Routing decisions are made internally by the provider, and the client only observes the resulting prompt-response pairs.
Without monetary compensation, this information gap creates an economic failure known as \emph{moral hazard}, as the provider can reduce costs by always using the low-cost routing policy internally. More formally, denote by $\vg_{1:t}\in\gG^t$ the hidden sequence of generative models used up to time $t$, and note that $\vg_{1:t}$ is random and depends on the choice of $\pi$. The cumulative cost of generating responses using these models is $c(\vg_{1:t})=\sum_{i=1}^t c_{g_i}$.
When high-effort routing is more costly
$\mathbb E_{\vg\sim \pi_\mathrm H}[c(\vg_{1:t})] > \mathbb E_{\vg' \sim \pi_\mathrm L}[c(\vg'_{1:t})]$, and without additional compensation,
a strategic provider knowing that contractual relations will terminate at time $t$ will deploy $\pi_\mathrm L$. We note that the expectation integrates over both the decisions made by the router and the randomness over prompts.

\paragraph{Pay for performance.}
To incentivize high-effort routing, the client commits to pay the provider based on measured performance. Formally, we assume
the existence of
a \emph{quality evaluation function}, denoted by $q$, which maps a prompt-response pair $(x,y)$ to a quality signal $s=q(x,y)$. We denote the space of signals by $\gS$ and the signal observed at time $t$ by $s_t\in\gS$.
The client commits to a contract which compensates the provider by transferring
$w(\vs_{1:t}) \geq 0$ monetary units upon termination, where $\vs_{1:t}$ is the observed signal
history.
The assumption of payment at contract termination is made for simplicity, and we note that other schedules such as monthly or per-use payment can be implemented via discounting and refund mechanisms.

Given contract $w(\vs_{1:t})$, the provider's utility when relations terminate at
time $t$ after deploying models $\vg_{1:t}$ and observing signals $\vs_{1:t}$ is thus:
\begin{equation}
\label{eq:fixed_time_util}
u(w; \vg_{1:t}, \vs_{1:t}) = w(\vs_{1:t}) - c(\vg_{1:t})
\end{equation}
Given knowledge of termination at time $t$ and now with commitment to receive compensation, the provider will deploy $\pi_\mathrm H$ if $\mathbb{E}_{\pi_\mathrm{H}}[u(w;\vg_{1:t}, \vs_{1:t})] \ge \mathbb{E}_{\pi_\mathrm{L}} [u(w;\vg'_{1:t}, \vs'_{1:t})]$.

Extending standard contract design, we consider contracts that contain an \emph{exit criterion}, allowing contractual relations to be terminated after any step and for any reason. Formally, let~$\gT$ be a subset of the set of integrable stopping times with respect to both $\pi_{\mathrm{H}}, \pi_\mathrm{L}$, for the filtration $\sigma(s_1, ..., s_t)$\footnote{Formally, a stopping rule is a function from observed sequences to scalars, such that $\tau(\vs_{1:t})=t$ if and only if $\tau$ stops the process at time $t$ after observing signals $\vs_{1:t}$.}. For example, a stopping rule $\tau\in\gT$ could terminate the contract after observing $\ell$ low-quality outputs, or after $T$ steps. Particularly notable instances are stopping times with uniformly bounded mean under both laws, and stopping times uniformly bounded by $T$, which we denote by~$\gT_T$. For a general stopping rule $\tau\in\gT$, termination time itself becomes a random variable,
and the expected utility of the provider given $\tau$ and payment rule $w$ is:
\begin{equation}
    U_\pi\left[w;\tau\right] =
    \mathbb E_{\pi}\left[
    u(w; \vg_{1:\tau}, \vs_{1:\tau})
    \right]
\end{equation}
In the following section, we show that incorporating exit clauses into contracts invalidates many of the existing approaches to contract design, which in turn motivates our construction of anytime contracts.

\section{Implications of stopping}
\label{sec:bernoulli}

In this section we show that moving from a fixed contract
duration, as in the classic setting, to an uncertain one has significant implications even when the underlying process is relatively simple.
In particular, we analyze contractual settings induced by \emph{independent Bernoulli processes}, in which quality signals are pass/fail, independent, and identically distributed --
capturing
settings where
outputs are considered sufficiently good if they meet
a threshold.
We present natural extensions of fixed-duration contract theory to indefinite durations, and show that they
fail to incentivize $\pi_\mathrm H$ once the contract can be adversarially terminated based on past performance.

 To begin, consider a setting in which the provider operates two models, denoted by $\gG=\Set{\mathrm{L}, \mathrm{H}}$, and the client seeks outputs generated by $\mathrm H$. Serving
model~$\mathrm H$
costs
$c_{\mathrm H}$ per step and returns a passing output with probability $p$;
model~$\mathrm L$
costs $c_{\mathrm L}$ per step and passes with
probability $q$. We assume $0<q<p<1$ and $0\le c_{\mathrm L}<c_{\mathrm H}$,
so that the model requested by the client has both higher quality and higher cost.
The high-quality router $\pi_\mathrm H$ always serves the high-quality model $\mathrm{H}$ and the deviation $\pi_\mathrm L$ always serves the low-quality one.
Signals are therefore i.i.d.\ $s_t\sim\mathrm{Bernoulli}(p)$ under $\pi_\mathrm H$, and
i.i.d.\ $s_t\sim\mathrm{Bernoulli}(q)$ under $\pi_\mathrm L$.

We now present two families of practically-motivated contracts, and analyze them from this perspective:

\paragraph{Failures of repeated optimal single-step contracts.}\label{ex:per-success}
A first attempt at designing contracts would be to simply repeatedly implement an optimal single-step contract. Such contracts typically pay out as a function of performance, for example by paying per high-quality output. Formally, consider $w_t(\vs_{1:t})=\alpha\sum_{i\le t}\mathbbm 1\{\vs_i=1\}$ for some $\alpha > 0$.
We note that such contracts are step-wise optimal, as optimal single-step contracts for Bernoulli outcomes only pay on success (e.g., \citet{saig2024incentivizing}, Prop.~3). We show that for all contracts of this type, if the ratio of success rates is smaller than the ratio of costs, under the stopping rule $\tau_{\mathrm{pass}} :=\inf\{t:s_t=1\}$ which represents clients that stop after a task is completed, the provider has higher utility if a low-quality model is deployed.
\begin{proposition}
    In Bernoulli contractual settings with $p c_\mathrm L < q c_\mathrm H$ and any step-wise contract, the provider benefits from using a low-quality policy under $\tau_{\mathrm{pass}}$.
\end{proposition}
Proof in Appendix~\ref{pf:per-success}.
We note that the conditions are often met in current practical settings. For example, in the datasets we analyze in Section~\ref{sec:real_data}, in particular Holistic
Agent Leaderboard \citep[HAL;][]{kapoor2026hal}, taking Claude Opus 4.1 as a ``high" model and Claude-3.7 Sonnet as a ``low" one yields $p/q = 0.61/0.5=1.22$ and $c_{\mathrm{H}} / c_{\mathrm{L}} = 27.03/8.05 \approx 3.36$. Moreover, this condition is met in any setting where the alternative model $\mathrm L$ has negligible cost $c_{\mathrm{L}} \approx 0$.

\paragraph{Failures of optimal fixed-duration contracts.}\label{ex:opt-fixed-horizon}
\label{ex:fixed-horizon-p4p} Another naive approach would be to design an optimal contract for a fixed duration. As we show, this approach also fails under a simple stopping rule.
For Bernoulli outcomes and a fixed duration $T$, the number of successful evaluations up to termination is a Binomial variable, and the fixed-duration contract minimizing expected pay
is the all-success rule $w_T(\vs_{1:T})=\alpha \mathbbm 1\{\vs_1=\dots=\vs_T=1\}$ with a suitable $\alpha\ge 0$ (e.g., \citet{dutting2019simple}, Lemma~7). We extend this
to any duration by taking $w(\vs_{1:t})=\alpha_t \mathbbm 1\{\vs_1=\dots=\vs_t=1\}$ at
whatever time the contract ends, and possibly tuning $\alpha_t$ for any $t$.
We show that under this family of contracts, the provider prefers $\pi_\mathrm L$ when $ \tau_{\mathrm{fail}}:=\inf\{t:s_t=0\}$, which represents clients that switch to a different provider after a failure:
\begin{proposition}
For any Bernoulli contractual setting and all-success pay, the provider benefits from using the low-quality policy under $\tau_{\mathrm{fail}}$.
\end{proposition}
Proof in \Cref{pf:opt-fix-horizon}. Intuitively, by construction there is one failure at stopping, and therefore
all-success payment is always zero, and the provider has incentive to deploy $\pi_\mathrm L$ due to lower operational costs.

Together, these examples indicate that a more suitable framework for this setting is required. Furthermore, we note that examples of stopping times that break incentive compatibility for simple contracts also exist for bounded stopping times -- See \Cref{pf:restrictive-spaces}.

\section{Anytime Contracts}
\label{sec:anytime_optimal_contracts}

Our results in the previous section highlight the need for accounting for stopping-time uncertainty in contract design. Since the stopping rule is in general private for the client organization and thus unknown to the provider, we study the problem of designing contracts that are robust to choices of stopping rules. As we show, this property gives rise to a new class of \emph{anytime contracts} which have natural connections with anytime-valid statistical tests.

\paragraph{Contracts with stopping-time robustness.}
In settings with stopping time uncertainty, the goal of the client organization is to design an efficient payment rule incentivizing the deployment of the high-effort routing policy $\pi_\mathrm H$ under any choice of stopping rule $\tau\in\gT$. Formally,
associating economic efficiency with required budget --- defined as the maximum theoretical transfer $\sup_{s, t}w(\vs_{1:t})$ --- contract design solves the following optimization problem:
\begin{equation}
\label{eq:general_optimization_problem}
\begin{aligned}
\argmin_{
w:\gS^+\to\R_{\ge0}
} \quad &\sup_{\vs,t}
w(\vs_{1:t}) \\
\text{ such that}\quad&
U_{\pi_\mathrm H}[w;\tau] \ge U_{\pi_\mathrm L}[w;\tau] &\forall \tau\in\gT.
\end{aligned}
\end{equation}
Consistent with contract design nomenclature, we refer to the constraints of Eq.~(\ref{eq:general_optimization_problem}) as \emph{incentive compatibility} constraints. When a contract $w$ satisfies these constraints, we say that it is an \emph{anytime contract} that implements $\pi_\mathrm H$.

 Our characterization has a natural formulation in terms of distributions over stopped sequences $\vs_{1:\tau}$, which we now formally define.
Denote by
$\gP_\pi(\cdot\mid\vs_{1:t})$
a version of the conditional distribution of the quality signal $s_{t+1}$ given the history up to that time.
We assume absolute continuity\footnote{The assumption of absolute-continuity  is made for convenience. If it fails, there are events that occur under \(\pi_{\mathrm L}\) but not under \(\pi_{\mathrm H}\); on such events the deviation is identified with certainty, so it is optimal to set the payment to zero. The remaining part of the problem is then treated on the absolutely continuous component.}
$
\gP_{\pi_\mathrm{L}} (\cdot \mid \vs_{1:t})
\ll
\gP_{\pi_\mathrm{H}} (\cdot \mid \vs_{1:t})
$
for every
$t, \vs_{1:t}$ and every history reachable under $\pi_{\mathrm L}$, so that the Radon-Nikodym derivative
$\mathrm{d}\gP_{\pi_\mathrm{L}}/\mathrm{d}\gP_{\pi_\mathrm{H}}$ is well-defined, and
we can therefore denote the likelihood ratio by $M(\vs_{1:t})
=\prod_{i=1}^{t}
\tfrac{\mathrm{d}\gP_{\pi_\mathrm{L}}}{\mathrm{d}\gP_{\pi_\mathrm{H}}}
(s_i | \vs_{1:{i-1}})$.
For a stopping rule $\tau\in\gT$, denote by $\gQ_\pi^\tau (\vs_{1:t})$ the probability of observing sequence $\vs_{1:t}$ when the contract terminates according to $\tau$. We note that $\gQ_\pi^\tau$ is defined over the set of nonempty signal sequences $\gS^+=\bigcup_{i=1}^\infty \gS^i$.

\paragraph{Budget-optimal anytime contracts.}
Using the notations defined above, Eq.~(\ref{eq:general_optimization_problem}) can be formulated as a linear program, in a possibly infinite-dimensional space and with an infinite number of constraints (see Appendix~\ref{app:lp}). The linear program is finite when $\gS$ is finite and stopping times are bounded, however we also note that even in the simple case of Bernoulli processes (defined in \Cref{sec:bernoulli}) and bounded stopping times $\gT_T$, the number of variables is exponential in $T$, and the number of constraints is doubly-exponential in $T$. Thus, direct naive optimization is computationally infeasible for all but very small signal spaces and values of $T$, further motivating the search for structure.
Our main result characterizes the solution of Eq.~(\ref{eq:general_optimization_problem}) in the general case.
\begin{theorem}
\label{thm:min-budget}
A budget-optimal contract implementing $\pi_\mathrm H$ under stopping-time uncertainty $\gT$ is:
\begin{equation}
\label{eq:opt_lr_contract}
w\left(\vs_{1:t}\right)=B^\star \cdot\indicator{M\left(\vs_{1:t}\right)\le 1}
\end{equation}
with required budget given by:
\begin{equation}
\label{eq:opt_lr_budget}
B^\star
=
\sup_{\tau\in\mathcal T}
\left(
  \frac
{U_{\pi_\mathrm L}[0;\tau]-U_{\pi_\mathrm H}[0;\tau]}
{\tv{\gQ^\tau_{\pi_\mathrm L}}{\gQ^\tau_{\pi_\mathrm H}}}
\right)^+
\end{equation}
where $M(\vs_{1:t})$ is the likelihood ratio at termination, $U_{\pi}[0;\tau]$ is the utility of the provider without any compensation under policy $\pi$, $\tv{\cdot}{\cdot}$ is the total variation distance, and $(\cdot)^+$ is the positive part.
\end{theorem}
Proof in Appendix~\ref{pf:min-budget}. The proof gives a direct measure-theoretic characterization of minimum-budget contracts that avoids the linear-program duality used in prior finite-horizon results~\citep[Thm.~1]{saig2023delegated}, allowing generalizations to settings with unbounded integrable stochastic stopping and possibly infinite signal spaces.
In addition, we note that the generalized min-budget characterization in our proof also expands the characterization frontier in non-robust settings, and may be of independent interest.

\paragraph{Statistical interpretation.}
The threshold rule in the optimal contract,
$\indicator{M(\vs_{1:t})\le1}$, also appears naturally in
hypothesis testing. More generally, the likelihood-ratio test
between $\pi_{\mathrm H}$ and $\pi_{\mathrm L}$ accepts the
null when $M(\vs_{1:t})\le1/\alpha$. When the test may be
evaluated at an arbitrary stopping time, $M_t$ is a
nonnegative martingale under $\pi_{\mathrm H}$, hence an
e-process, and the resulting test is anytime-valid.
Likelihood-ratio e-processes are also optimal under
log-growth criteria in sequential testing
\citep[Sec.~3.5]{ramdas2025hypothesis}. Thus the same
likelihood-ratio statistic arises from two different
optimality criteria: growth in anytime-valid testing and
budget in our contract design problem. The parameter $\alpha$ controls how protective the audit is
for the provider. Ville's inequality gives
$\mathbb P_{\pi_{\mathrm H}}
(M(\vs_{1:\tau})>1/\alpha)\le\alpha$
for every $\tau\in\gT$. Thus a provider deploying
$\pi_{\mathrm H}$ is denied payment with probability at most
$\alpha$, uniformly over stopping rules. The budget-optimal
contract corresponds to the strictest threshold of $\alpha=1$.

\subsection{Computing Anytime Contracts}\label{sec:algorithms}

We now compute the contract of Theorem~\ref{thm:min-budget} in the setting where $\gT = \gT_T$ and a finite signal space $\mathcal{S}$. The optimal budget is a supremum over stopping times, of which there are doubly exponentially many in $T$, hence a naive enumeration does not work. However, extracting a worst-case stopping rule for a candidate budget is a dynamic programming problem. Finding a minimal such budget then reduces to a finiteness check and a binary search over the budget parameter.

For a history $h=\vs_{1:t}$ and $\alpha \in (0, 1]$, denote $w_B(h) = B\,\indicator{M(h) \leq 1/\alpha}$ and $\widehat c_{\pi}(h) = \mathbb E_\pi[c(\vg_{1:t})\mid\vs_{1:t}=h]$, and let the stopping reward be $R_B(h)=\widehat c_{\pi_{\mathrm H}}(h)-M(h)\,\widehat c_{\pi_{\mathrm L}}(h)-\bigl(1-M(h)\bigr)\,w_B(h)$. Dynamic programming applies because of the structure of stopping rules: at every history, a stopping rule either stops and collects the reward $R_B(h)$, or continues to the next signal, where it faces the same choice~\citep{peskir2006optimal}. Therefore $V_B(h):=\max_{\tau\in\gT_T,\,\tau\ge|h|}\mathbb E_{\pi_{\mathrm H}}[R_B(\vs_{1:\tau})\mid\vs_{1:|h|}=h]$ satisfies the Bellman recursion:
\begin{equation}
\label{eq:bellman-recursion-main}
V_B(h)=\max\Bigl\{R_B(h),\ \textstyle\sum_s\gP_{\pi_{\mathrm H}}(s\mid h)\,V_B(h,s)\Bigr\},
\end{equation}
with $V_B(h)=R_B(h)$ for $|h|=T$. A single backward pass over the tree of reachable histories, whose size we denote by $n$, computes $V_B(\emptyset)$, the largest incentive violation of $w_B$ over $\gT_T$. We call this pass \textsc{Verify}$(B)$, so that $w_B$ is incentive compatible exactly when \textsc{Verify}$(B)\le0$. Since \textsc{Verify}$(B)$ is nonincreasing in $B$, the minimal budget $B_\alpha^\star$ is found by binary search after a finiteness check.
The following theorem formalizes this approach, details are in Appendix~\ref{app:algorithms}:

\begin{theorem}[Correctness and complexity]
\label{thm:compute-budget}
Given a history representation of size $n$,
\textsc{Verify}$(B)$ computes the largest incentive violation
in $O(n)$ arithmetic operations and oracle calls. Finiteness
of $B_\alpha^\star$ can also be checked in $O(n)$ operations.
If $B_\alpha^\star<\infty$, the resulting search returns an
incentive-compatible $w_U$ satisfying
$B_\alpha^\star\le U\le B_\alpha^\star+\varepsilon$ using
$O\!\left(n\log\!\left(\frac{B_\alpha^\star}{\varepsilon}\right)\right)$
arithmetic operations and oracle calls.
\end{theorem}

The full history tree has $n=O(|\gS|^T)$ nodes. However, since $w_B$ depends on $h$ only through $M(h)$, whenever $M$, $\widehat c_\pi$ and $\gP_{\pi_{\mathrm H}}$ depend on $h$ only through its signal counts, histories with equal counts can be merged, giving $n=O(T^{|\gS|})$ and computation polynomial in $T$. For i.i.d.\ Bernoulli setting of Section~\ref{sec:bernoulli}, this is $O(T^2)$.

\begin{figure*}[ht!]
    \includegraphics[width=\textwidth]{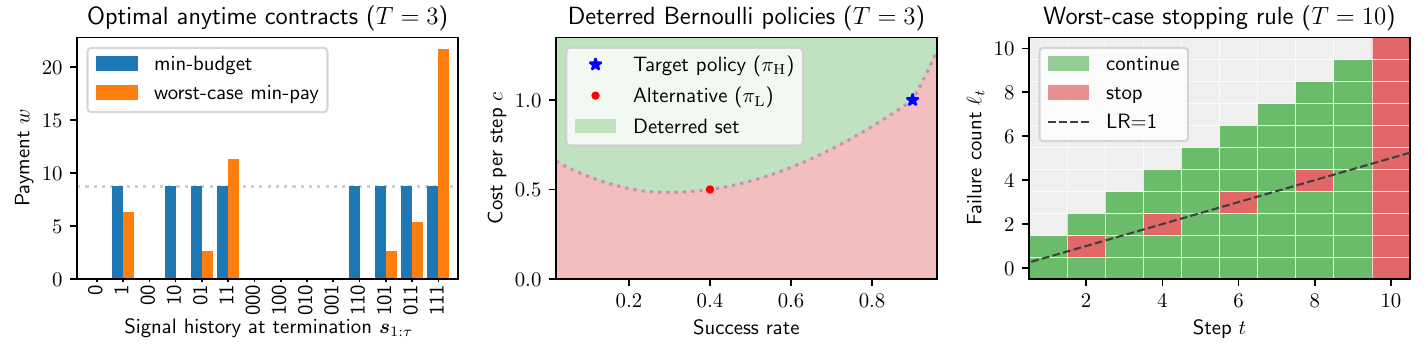}
    \caption{Anytime contracts
    for synthetic Bernoulli processes
    (\Cref{sec:synthetic_data}).
    \textbf{(Left)} Optimal anytime contracts under different objectives.
    \textbf{(Center)} Non-monotone deterred set.
    \textbf{(Right)} Worst-case stopping rule.
    }
    \label{fig:synthetic}
\end{figure*}

\section{Partial Prior Knowledge}\label{sec:incomplete}

The optimal contract of
Section~\ref{sec:anytime_optimal_contracts} has a simple form,
but it may be difficult to implement due to the fact that its \emph{audit} mechanism---captured in the condition $\mathbbm 1\{M_t\le1\}$---requires knowledge of the likelihood ratio $M_t$,
and hence knowledge of the distributions of the signals induced by both routing
policies. In most settings, the client organization or third-party evaluator
lacks direct knowledge of the underlying models and only observes the provider's evaluation history. In this section, we construct a principled audit directly from the observed signal history; so that the provider's performance can be audited \emph{without the auditor's knowledge of the underlying models}.

We assume that the signal space
consists of bounded scalar evaluations
$\gS\subseteq[a,b]\subset\mathbb R$, and construct the audit using the testing-by-betting framework of
\citet{waudbysmith2024}. The resulting contract inherits statistical properties: If the provider commits to
maintaining a reference quality level $s_0\in(a,b)$, the same audit gives
an anytime-valid level-$\alpha$ guarantee against denial of
payment. Moreover, when the mean quality of $\pi_\mathrm H$ is $s_0$, every audit in the constructed family is exactly a
likelihood-ratio audit against an associated alternative
signal law.

To construct the audit, fix a reference level $s_0\in(a,b)$, and initialize $E_0=1$.
Before observing $s_t$ the auditor chooses a nonnegative stake
$\lambda_t$, which may depend on the previous signals but not
on $s_t$, and updates
$
E_t
=
E_{t-1}\bigl(1+\lambda_t(s_0-s_t)\bigr).
$
We require
$0\le\lambda_t\le1/(b-s_0)$, ensuring that the process remains
nonnegative. The interpretation is that $E_t$ accumulates evidence against
performance at the reference level: it increases when
$s_t<s_0$ and decreases when $s_t>s_0$. For $\alpha\in(0,1]$, we define the audit
$\phi(\vs_{1:t}) =\mathbbm 1\{E_t\le1/\alpha\}$ and the corresponding
contract $w_{\alpha, B}^{\mathrm{bet}}=B\phi_t$. Once $s_0$ and the
stake rule are specified, the audit can be evaluated
sequentially from the observed signals alone.

\paragraph{Service-level protection.}
Suppose the provider guarantees service level
$
\mathbb E_{\pi_{\mathrm H}}
[s_t\mid\vs_{1:t-1}]
\ge s_0
$
at every reachable history. Because $\lambda_t$ is chosen
before $s_t$,
$
\mathbb E_{\pi_{\mathrm H}}
\!\left[
1+\lambda_t(s_0-s_t)
\mid\vs_{1:t-1}
\right]
\le1,
$
so $E_t$ is a nonnegative supermartingale. Ville's
inequality therefore implies
$
\inf_{\tau\in\gT}
\mathbb P_{\pi_{\mathrm H}}
(E_\tau\le1/\alpha)
\ge1-\alpha
$.
Thus a provider satisfying the advertised service level $s_0$
receives level-$\alpha$ protection against denial of payment,
uniformly over when the contract is stopped -- See \Cref{app:bet_service}.

\paragraph{Relation to the likelihood-ratio audit.}
Suppose now that the promised router lies exactly on the service-level
boundary,
$
\mathbb E_{\pi_{\mathrm H}}
[s_t\mid\vs_{1:t-1}]
=
s_0.
$
For any admissible predictable stake $\lambda_t$, define an
associated conditional signal law by
$
\gP_t^\lambda(ds\mid h)
=
\bigl(1+\lambda_t(h)(s_0-s)\bigr)
\gP_{\pi_{\mathrm H}}(ds\mid h).
$
The conditional-mean identity ensures that this is a
probability kernel. Its one-step likelihood ratio relative to
the promised law is exactly
$1+\lambda_t(h)(s_0-s)$, and therefore its running likelihood
ratio is
$
M_t^\lambda
=
E_t
$ -- See \Cref{app:bet_lr}.

Finally, we note that an analogous algorithmic approach as described in Section~\ref{sec:algorithms} can be used to find the minimal budget that makes $w_{\alpha, B}^{\mathrm{bet}}$ incentive compatible in the class $\gT_T$ and assuming a finite signal space $\mathcal{S}$. We relegate the details to the Appendix~\ref{app:bet_pf}.

\section{Empirical Evaluation}\label{sec:experiments}

\begin{figure*}[ht!]
    \includegraphics[width=\textwidth]{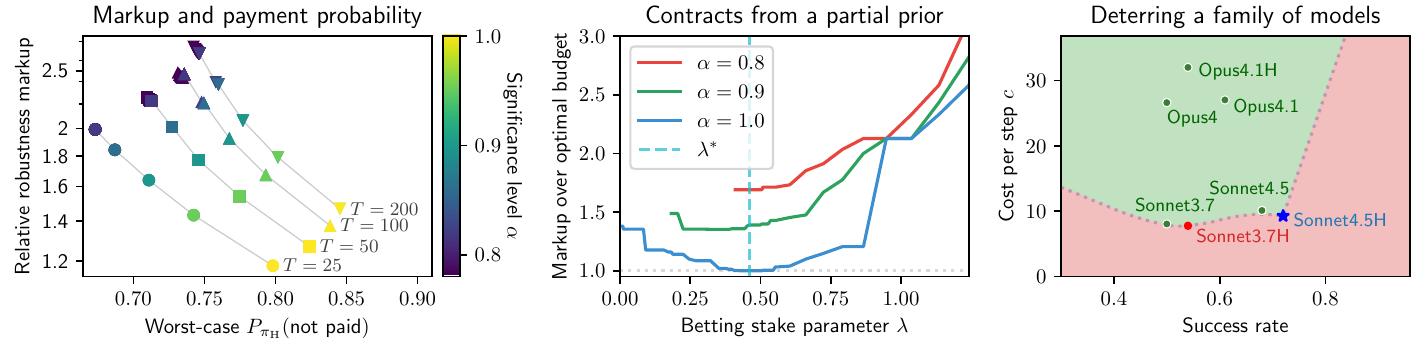}
    \caption{Empirical contracts based on the HAL dataset (\Cref{sec:real_data}).
    \textbf{(Left)} Robustness markup and probability of no monetary transfer, for varying significance levels and horizons.
    \textbf{(Center)} Betting e-value contracts from first-moment information.
    \textbf{(Right)} Deterring a family of alternative models with a two-action contract.
    }
    \label{fig:hal}
\end{figure*}

In this section, we develop intuition about anytime contracts and evaluate them in empirical settings.
Throughout, the evaluation signal is Bernoulli, as described in \Cref{sec:bernoulli}. Exit times $\gT_T$ are uniformly bounded, costs are per step, and we use the algorithm presented in \Cref{sec:algorithms} for computation. Constants and implementation details are given in  \Cref{app:reproduction}.

\subsection{Qualitative Analysis}\label{sec:synthetic_data}
We begin by qualitatively analyzing the structure imposed by optimal contracts on synthetic examples. We do so by comparing min-budget and min-pay, investigating the set of routers a contract deters, and visualizing worst-case stopping rules.

\paragraph{Min-budget and min-pay.}
Figure~\ref{fig:synthetic} (Left) compares the min-budget and worst-case min-pay
objectives in a Bernoulli instance.
The min-pay optimum,
obtained via solving the linear program in Appendix~\ref{app:lp}, has a qualitatively different structure from the min-budget one, with payments unequal across histories. This contrasts the fixed-duration setting,
where contracts optimal under both objectives can be expressed as scaled
statistical tests~\citep[Thm.~1]{saig2024incentivizing}, indicating that
a foundational discrepancy between the two objectives arises from adversarial stopping robustness.

\paragraph{Deterred set.}
A contract designed to incentivize one router against another may deter
others too, but not all of them. For Bernoulli signals, we design a contract incentivizing $(p,c_{\mathrm H})$ against $(q,c_{\mathrm L})$, and measure incentive compatibility for other success-cost pairs $(q',c')$. \Cref{fig:synthetic} (Center) outlines the results. We observe that the boundary is not monotone in $q'$, so incentive compatibility against $(q,c_{\mathrm L})$ does not guarantee deterrence of a policy  $(q',c')$ that is both less accurate
and more expensive.
This contrasts fixed-duration settings, where success counts are Binomial, and such monotonicity holds due to their monotone likelihood ratio property (e.g., by \citep[Lem. 3.4.2]{lehmann2005testing}).

\paragraph{Worst-case exit rules.}
\Cref{fig:synthetic} (Right) displays the worst-case exit rule returned by the dynamic programming algorithm of \Cref{sec:algorithms}. The organization stops where the
likelihood ratio leaves the audit's verdict the closest to the threshold of 1. As a possible interpretation, when the ratio is
small or large the verdict is already anticipated, so the
worst case stopping rule keeps consuming to drain the
provider's profit.

\subsection{Practical Considerations}\label{sec:real_data}

We now calibrate the same objects on deployed models, using empirical data from the Holistic
Agent Leaderboard \citep[HAL;][]{kapoor2026hal} to set $p$, $q$,
$c_{\mathrm H}$ and $c_{\mathrm L}$. Accuracies and costs are the SWE-Agent
entries of the SWE-bench Verified Mini leaderboard
\citep{hal2026swebenchmini}, a $50$-task subset of the SWE-bench Verified benchmark
\citep{jimenez2024swebench}. We evaluate three economic quantities: the role of $\alpha$ in guaranteeing providers are paid more often, the price of not knowing the observation laws, and whether a
single contract can deter an entire family of models.

\paragraph{The role of $\alpha$.}
Figure~\ref{fig:hal} (Left) traces the implementable $\alpha$ range for Claude Opus 4.1 over
Claude-3.7 Sonnet. The vertical axis is the \emph{price
of robustness} $w^\star_\alpha/B^{\mathrm{fix}}_{T, \alpha}$, the budget
measured against the one that would suffice if the exit date were known to
be $T$; the horizontal axis is the worst-case
probability that an honest provider is denied payment. Lowering $\alpha$
raises the cutoff $1/\alpha$, so the audit pays on more histories: the price
rises and the denial probability falls. The knob is thus a dial for favoring
the provider, priced accordingly. Smaller values of $\alpha$ need a larger markup as the additional slack given to the provider increases information gaps. Larger $T$
lifts every arc, since a longer duration supplies more
exit rules to choose from and the supremum runs over a larger class.

\paragraph{The price of partial information.}
Keeping the same pair, Figure~\ref{fig:hal} (Center) reports the wage of the
betting audit of Section~\ref{sec:incomplete} relative to the
full-information min-budget contract of Theorem~\ref{thm:min-budget}, as a
function of the stake $\lambda$, and with the announced service level set to $s_0=p$. At $\lambda^\star=(1-q/p)/(1-p)$ the betting process
recovers the likelihood ratio as discussed in Section~\ref{sec:incomplete}, achieving a global minimum markup of 1. The approximation is forgiving near
$\lambda^\star$, allowing for near optimal audits even with partial information.

\paragraph{Deterring a family of models.}
Finally, we revisit the deterred set with empirical data.
\Cref{fig:hal} (Right) outlines the deterrence set for the Claude family of models, induced by a contract targeting Sonnet 4.5H, the highest-performing model in the family appearing in the leaderboard, against Sonnet 3.7H as the reference alternative. Despite being designed against a single reference model, the contract incentivizes the target model over every alternative Claude model appearing on the leaderboard, even though the resulting deterrence set is neither monotone nor convex.

\section{Conclusion \& limitations}\label{sec:conclusion}

In this work,
we introduce a contract-theoretic framework that explicitly accounts for stopping-time uncertainty and addresses the moral hazard problem inherent in AI services. We show that several natural constructions from the fixed-horizon setting fail under this stronger requirement, and characterize a budget-optimal contract that remains incentive compatible across admissible stopping rules. The resulting contract is efficiently computable via dynamic programming, enabling empirical analysis on both synthetic instances and data from deployed models.

Our framework also relies on several assumptions that motivate further inquiry. In particular, we assume that the client has sufficient bargaining power, or can coordinate with others to exercise such power; model the principal--agent interaction in isolation from other market forces; assume that clients are likely to exercise the most adversarial stopping rule in $\gT$; and require stopping rules to be integrable. Relaxing these assumptions raises intriguing questions for future work.

\subsection*{Acknowledgements}
Eden Saig is supported in part by the Zuckerman STEM Leadership Program, and by the AISI Alignment Project.

\subsection*{AI Use Statement}
In this work, we used generative AI tools to assist in proving our theoretical results, to help formulate the linear-programming formulation in Appendix~\ref{app:lp}, to generate code implementing our methods and experiments, and to obtain feedback on drafts of the paper, including on the research methodology and experiments. We have not used generative AI tools to formalize the problem, to develop research ideas or hypotheses, or to interpret our results. Generating or cleaning data sets, translation, and qualitative data analysis are not applicable to this work.
Additionally, we used generative AI tools to identify relevant literature, to format references, and to improve the grammar, flow, and readability of the text and the typesetting of mathematical formulas; no sections of the paper were drafted by generative AI tools.

We have reviewed all AI-assisted work: all AI-assisted proofs and derivations were checked line by line by the authors; all AI-generated code was read and reviewed by the authors and tested on independent small-scale instances, where we verified that independent solvers return the same solutions; all suggested references were verified against the original sources; and all AI-edited text was read and revised by the authors. We take responsibility for the final content of this work, including text, claims, or artifacts produced with the aid of generative AI.

\bibliography{citations}
\bibliographystyle{plainnat}

\clearpage
\appendix
\thispagestyle{empty}

\onecolumn
\papertitle{Anytime Contracts for AI Services: \\
Supplementary Materials}

\crefalias{section}{appendix}
\crefalias{subsection}{subappendix}
\crefalias{subsubsection}{subsubappendix}

\section{Implications of stopping}\label{sec:naive}

\subsection{Payments depending only on time.}\label{app:pay_time}
In this family of contracts, the client's payment is only dependent on time; formally $w(\vs_{1:t})=\gamma(t)$ for some increasing function $\gamma$, e.g. fixed payment per step $\gamma(t)=\alpha t$. Assuming that a deterministic stopping time $\tau = T$ is in $\gT$, we immediately obtain that the contract will not be incentive compatible for that $\tau$. Formally, the provider utility is
$U_{\pi}[w;\tau] = \gamma(T) - c_{\pi}T$,
and as $c_{\mathrm{H}} > c_{\mathrm{L}}$ we obtain $U_{\pi_{\mathrm{H}}}[w;\tau] < U_{\pi_{\mathrm{L}}}[w;\tau]$. Hence the low-quality policy is always beneficial for the provider in such cases.

\subsection{Step-wise pay-for-performance.}\label{pf:per-success}

By construction $\vs_1=\dots=\vs_{\tau_{\mathrm{pass}}-1}=0$ and
$\vs_{\tau_{\mathrm{pass}}}=1$, so the running count of successes equals one on every path, and $w(\vs_{1:\tau_{\mathrm{pass}}})=\alpha$ almost surely under both routers. Hence the utilities are:
$$
 E_{\pi_{\mathrm{H}}}[w(\vs_{1:\tau_{\mathrm{pass}}}) - c_{\mathrm{H}}\tau_{\mathrm{pass}}]=\alpha-\frac{c_{\mathrm H}}{p}, \quad E_{\pi_{\mathrm{L}}}[w(\vs_{1:\tau_{\mathrm{pass}}}) - c_{\mathrm{L}}\tau_{\mathrm{pass}}]=\alpha-\frac{c_{\mathrm L}}{q},
$$
Stopping time $\tau_{\text{pass}}$ satisfies $U_{\pi_\mathrm{H}}[w; \tau_{\mathrm{pass}}] \geq U_{\pi_\mathrm{L}}[w; \tau_{\mathrm{pass}}]$ if and only if $p / q \geq c_{\mathrm H} / c_{\mathrm L}$. Somewhat surprisingly, even for arbitrarily large $\alpha$, the contract may fail to be incentive compatible, as the constraint for $\tau_{\mathrm{pass}}$ is wage-agnostic.

\subsection{Optimal fixed-horizon contracts}\label{pf:opt-fix-horizon}

By construction $s_{\tau_{\mathrm{fail}}} = 0$, therefore $w(\vs_{1:\tau_{\mathrm{fail}}}) = 0$. The utilities therefore are
$$
 E_{\pi_{\mathrm{H}}}[w(\vs_{1:\tau_{\mathrm{fail}}}) - c_{\mathrm{H}}\tau_{\mathrm{fail}}]=-\frac{c_{\mathrm H}}{1-p}, \quad E_{\pi_{\mathrm{L}}}[w(\vs_{1:\tau_{\mathrm{fail}}}) - c_{\mathrm{L}}\tau_{\mathrm{fail}}]=-\frac{c_{\mathrm L}}{1-q},
$$
as $c_{\mathrm{H}} > c_{\mathrm{L}}, p > q$ it always holds that $U_{\pi_\mathrm{H}}[w; \tau_{\mathrm{fail}}] < U_{\pi_\mathrm{L}}[w; \tau_{\mathrm{fail}}]$.

\subsection{Failure of incentive compatibility in $\gT_T$}\label{pf:restrictive-spaces}
The exit rules of Section~\ref{sec:bernoulli} are unbounded. Restricting to
$\gT_T$ does not help. We now give such an instance.

Take $T=2$, $p=\tfrac34$, $q=\tfrac14$, $c_{\mathrm L}=0$, and the contract
$w_t(\vs_{1:t})=\alpha_t\,\mathbbm 1\{\vs_1=\dots=\vs_t=1\}$ with
$\alpha_t=c_{\mathrm H}t/(p^{\,t}-q^{\,t})$, so that
$\alpha_1=2c_{\mathrm H}$ and $\alpha_2=4c_{\mathrm H}$. Note that this is the closed-form solution that is min-pay optimal~\cite{dutting2019simple}. Therefore, the contract is incentive-compatible at each deterministic $\tau = t$. Now let the client leave on a success and otherwise run to the bound,
\[
\tau:=\tau_{\mathrm{pass}}\wedge T=
\begin{cases}
1, & \vs_1=1,\\[2pt]
2, & \vs_1=0,
\end{cases}
\]
a stopping time bounded by $T$, hence in $\gT_T$. A straightforward calculation gives us the utilities for ``high" and ``low" models, respectively:
$
\tfrac34\bigl(2c_{\mathrm H}-c_{\mathrm H}\bigr)+\tfrac14\bigl(-2c_{\mathrm H}\bigr)
=\tfrac14 c_{\mathrm H},
\tfrac14\bigl(2c_{\mathrm H}\bigr)=\tfrac12 c_{\mathrm H},
$
so the deviation is strictly better off for every $c_{\mathrm H}>0$.

\section{Anytime Contracts}

\subsection{Linear Programming Formulations}\label{app:lp}

Since $\mathbb E_\pi[w(\vs_{1:\tau})]=\int w\,\mathrm d\gQ^\tau_\pi$, incentive compatibility at $\tau$ is linear in $w$:
\begin{equation}
\label{eq:lp-ic}
\int_{\gS^+} w\,\mathrm d\bigl(\gQ^\tau_{\pi_\mathrm H}-\gQ^\tau_{\pi_\mathrm L}\bigr)\;\ge\;U_{\pi_\mathrm L}[0;\tau]-U_{\pi_\mathrm H}[0;\tau]
=\mathbb E_{\pi_\mathrm H}[c(\vg_{1:\tau})]-\mathbb E_{\pi_\mathrm L}[c(\vg_{1:\tau})].
\end{equation}
Introducing epigraph variables for the objectives, the min-budget and min-pay problems are the linear programs
\begin{align}
\label{eq:lp-budget}
&\min_{w\ge0,\,B}\ B
&&\text{s.t.}\quad \eqref{eq:lp-ic}\ \ \forall\tau\in\gT,
\qquad w(h)\le B\ \ \forall h\in\gS^+,\\
\label{eq:lp-pay}
&\min_{w\ge0,\,z}\ z
&&\text{s.t.}\quad \eqref{eq:lp-ic}\ \text{ and }\ \int_{\gS^+} w\,\mathrm d\gQ^\tau_{\pi_\mathrm H}\le z\ \ \forall\tau\in\gT,
\end{align}
over measurable $w:\gS^+\to\R_{\ge0}$. Both programs have one constraint per stopping rule, and are infinite-dimensional when $\gS$ is infinite or $\gT$ is unbounded.

\paragraph{Finite case.}
Suppose $\gS$ is finite and $\gT=\gT_T$, and let $\gH=\bigcup_{t=1}^T\gS^t$; payments on histories longer than $T$ are irrelevant. For $h\in\gS^t$, define
\[
\nu(h)=\mathbb P_{\pi_\mathrm H}(\vs_{1:t}=h),\qquad
\mu(h)=\mathbb P_{\pi_\mathrm L}(\vs_{1:t}=h),\qquad
C_\pi(h)=\mathbb E_\pi\bigl[c(\vg_{1:t})\,\indicator{\vs_{1:t}=h}\bigr].
\]
For adaptive routers, $\nu$ and $\mu$ are products of the conditional signal probabilities along $h$.
Identify each $\tau\in\gT_T$ with its stopping set $A_\tau\subseteq\gH$, the set of histories at which it stops, so that every history in $\gS^T$ has exactly one prefix in $A_\tau$. Then $\gQ^\tau_{\pi_\mathrm H}$ and $\gQ^\tau_{\pi_\mathrm L}$ coincide with $\nu$ and $\mu$ on $A_\tau$, and \eqref{eq:lp-ic} becomes
\[
\sum_{h\in A_\tau}\bigl(\nu(h)-\mu(h)\bigr)\,w(h)\;\ge\;\sum_{h\in A_\tau}\bigl(C_{\pi_\mathrm H}(h)-C_{\pi_\mathrm L}(h)\bigr),
\]
while the min-pay constraint becomes $\sum_{h\in A_\tau}\nu(h)\,w(h)\le z$. Both programs thus have $|\gH|+1$ variables. The number of stopping sets is $g_T^{|\gS|}$, where $g_1=1$ and $g_j=1+g_{j-1}^{|\gS|}$, which grows doubly exponentially in $T$ whenever $|\gS|\ge2$.

\subsection{Proof of Theorem~\ref{thm:min-budget}}\label{pf:min-budget}

In Eq.~\ref{eq:opt_lr_budget}, we interpret $(a/0)^+$ as $0$
when $a\le0$ and as $+\infty$ when $a>0$, and take
$\sup\emptyset=0$. The contract in Eq.~\ref{eq:opt_lr_contract}
is understood for $B^\star<\infty$; when $B^\star=\infty$, the
assertion is that no finite-budget contract implements
$\pi_{\mathrm H}$.

Note that for any nonnegative $\sigma(\vs_{1:t})$-measurable
function $Z_t$, by the definition of the likelihood ratio as the
Radon--Nikodym derivative, we have
\[
\mathbb{E}_{\pi_\mathrm{L}}[Z_t]
=
\mathbb{E}_{\pi_\mathrm{H}}[M(\vs_{1:t})Z_t],
\]
where the expectation is taken over the induced distribution of
signals by $\pi_\mathrm{H},\pi_\mathrm{L}$.

As $\tau\in\gT$ is an integrable stopping time with respect to both
$\pi_\mathrm{L},\pi_\mathrm{H}$, it is almost surely finite, and
the following decomposition holds
$\pi_\mathrm{L},\pi_\mathrm{H}$-almost surely:
\[
\begin{aligned}
w(\vs_{1:\tau})
&=
\sum_{t=1}^{\infty}
w(\vs_{1:t})\mathbbm{1}\{\tau=t\},\\
M(\vs_{1:\tau})w(\vs_{1:\tau})
&=
\sum_{t=1}^{\infty}
M(\vs_{1:t})w(\vs_{1:t})\mathbbm{1}\{\tau=t\}.
\end{aligned}
\]
Now note that $w(\vs_{1:t})\mathbbm{1}\{\tau=t\}$ is a
$\sigma(\vs_{1:t})$-measurable function, as $\tau$ is a stopping
time for this filtration. Therefore, utilizing Tonelli's theorem
and nonnegativity of $w,M$:
\begin{align*}
\mathbb{E}_{\pi_\mathrm{L}}[w(\vs_{1:\tau})]
&=
\mathbb{E}_{\pi_\mathrm{L}}\bigg[
\sum_{t=1}^{\infty}
w(\vs_{1:t})\mathbbm{1}\{\tau=t\}\bigg]\\
&=
\sum_{t=1}^{\infty}
\mathbb{E}_{\pi_\mathrm{L}}\big[
w(\vs_{1:t})\mathbbm{1}\{\tau=t\}\big]\\
&=
\sum_{t=1}^{\infty}
\mathbb{E}_{\pi_\mathrm{H}}\big[
M(\vs_{1:t})w(\vs_{1:t})\mathbbm{1}\{\tau=t\}\big]\\
&=
\mathbb{E}_{\pi_\mathrm{H}}\bigg[
\sum_{t=1}^{\infty}
M(\vs_{1:t})w(\vs_{1:t})\mathbbm{1}\{\tau=t\}\bigg]\\
&=
\mathbb{E}_{\pi_\mathrm{H}}\big[
M(\vs_{1:\tau})w(\vs_{1:\tau})\big].
\end{align*}
Since this identity holds for every nonnegative measurable
function of the stopped signal sequence,
$M(\vs_{1:\tau})$ is the Radon--Nikodym derivative of
$\gQ^\tau_{\pi_\mathrm{L}}$ with respect to
$\gQ^\tau_{\pi_\mathrm{H}}$.

Fix arbitrary $\tau\in\gT$ and an incentive-compatible $w$
with finite budget $\sup_{\vs,t}w(\vs_{1:t})<\infty$. In particular, for each
$\pi\in\{\pi_{\mathrm H},\pi_{\mathrm L}\}$,
\[
0\le
\mathbb E_\pi[w(\vs_{1:\tau})]
\le
\sup_{\vs,t}w(\vs_{1:t})
<
\infty.
\]
Moreover, since $\gG$ is finite and $\tau$ is integrable under
both routers,
\[
0\le
\mathbb{E}_{\pi}[c(\vg_{1:\tau})]
\le
\big(\max_{g\in\gG}c_g\big)\mathbb{E}_{\pi}[\tau]
<
\infty.
\]
Thus both expected payments and both expected costs are finite,
which justifies the subsequent rearrangements of the incentive
compatibility constraint.

Note that
\[
U_{\pi_\mathrm{L}}[0;\tau]
-
U_{\pi_\mathrm{H}}[0;\tau]
=
\mathbb{E}_{\pi_\mathrm{H}}[c(\vg_{1:\tau})]
-
\mathbb{E}_{\pi_\mathrm{L}}[c(\vg_{1:\tau})],
\]
so this is simply a compact way of referring to the cost
differential. Therefore, unpacking the incentive compatibility
constraint for $\tau$, we obtain that it is equivalent to
\[
\mathbb{E}_{\pi_\mathrm{H}}[w(\vs_{1:\tau})]
-
\mathbb{E}_{\pi_\mathrm{L}}[w(\vs_{1:\tau})]
\ge
U_{\pi_\mathrm{L}}[0;\tau]
-
U_{\pi_\mathrm{H}}[0;\tau].
\]

Consequently, using the change of measure and linearity of
expectation, the constraint is equivalent to
\[
\mathbb{E}_{\pi_\mathrm{H}}\big[
w(\vs_{1:\tau})(1-M(\vs_{1:\tau}))\big]
\ge
U_{\pi_\mathrm{L}}[0;\tau]
-
U_{\pi_\mathrm{H}}[0;\tau].
\]

Next, note that the following inequality holds pointwise:
\[
\begin{aligned}
w(\vs_{1:\tau})(1-M(\vs_{1:\tau}))
&\le
w(\vs_{1:\tau})(1-M(\vs_{1:\tau}))
\mathbbm{1}\{M(\vs_{1:\tau})\le1\}\\
&\le
\big(\sup_{\vs,t}w(\vs_{1:t})\big)
(1-M(\vs_{1:\tau}))
\mathbbm{1}\{M(\vs_{1:\tau})\le1\}.
\end{aligned}
\]
Taking expectations of both sides, we obtain
\[
\begin{aligned}
&\mathbb{E}_{\pi_\mathrm{H}}\big[
w(\vs_{1:\tau})(1-M(\vs_{1:\tau}))\big]\\
&\qquad\le
\big(\sup_{\vs,t}w(\vs_{1:t})\big)
\mathbb{E}_{\pi_\mathrm{H}}\big[
(1-M(\vs_{1:\tau}))
\mathbbm{1}\{M(\vs_{1:\tau})\le1\}\big]\\
&\qquad=
\big(\sup_{\vs,t}w(\vs_{1:t})\big)
\tv{\gQ^\tau_{\pi_\mathrm L}}{\gQ^\tau_{\pi_\mathrm H}},
\end{aligned}
\]
where the last equality is the total variation identity for
two probability measures and their Radon--Nikodym derivative.
Combining with incentive compatibility, we obtain
\[
\big(\sup_{\vs,t}w(\vs_{1:t})\big)
\tv{\gQ^\tau_{\pi_\mathrm L}}{\gQ^\tau_{\pi_\mathrm H}}
\ge
U_{\pi_\mathrm{L}}[0;\tau]
-
U_{\pi_\mathrm{H}}[0;\tau].
\]
If the total variation is zero and the cost differential is
positive, we obtain $B^\star = +\infty$ using our conventions, and this inequality rules out every finite-budget
contract. Otherwise, using nonnegativity of the budget and
the convention above, we obtain
\[
\sup_{\vs,t}w(\vs_{1:t})
\ge
\left(
\frac{U_{\pi_\mathrm{L}}[0;\tau]
      -U_{\pi_\mathrm{H}}[0;\tau]}
{\tv{\gQ^\tau_{\pi_\mathrm L}}{\gQ^\tau_{\pi_\mathrm H}}}
\right)^+.
\]
Taking the supremum over all $\tau\in\gT$, we achieve the
lower bound
\[
\sup_{\vs,t}w(\vs_{1:t})
\ge
\sup_{\tau\in\gT}
\left(
\frac{U_{\pi_\mathrm{L}}[0;\tau]
      -U_{\pi_\mathrm{H}}[0;\tau]}
{\tv{\gQ^\tau_{\pi_\mathrm L}}{\gQ^\tau_{\pi_\mathrm H}}}
\right)^+
=
B^\star.
\]

Suppose now that $B^\star<\infty$. It remains to show that this budget is
achieved. Define
\[
w(\vs_{1:t})
=
B^\star\mathbbm{1}\{M(\vs_{1:t})\le1\}.
\]
For every $\tau\in\gT$, we have
\[
\begin{aligned}
&\mathbb{E}_{\pi_\mathrm{H}}\big[
w(\vs_{1:\tau})(1-M(\vs_{1:\tau}))\big]\\
&\qquad=
B^\star
\mathbb{E}_{\pi_\mathrm{H}}\big[
(1-M(\vs_{1:\tau}))
\mathbbm{1}\{M(\vs_{1:\tau})\le1\}\big]\\
&\qquad=
B^\star
\tv{\gQ^\tau_{\pi_\mathrm L}}{\gQ^\tau_{\pi_\mathrm H}}\\
&\qquad\ge
\big(U_{\pi_\mathrm{L}}[0;\tau]
     -U_{\pi_\mathrm{H}}[0;\tau]\big)^+\\
&\qquad\ge
U_{\pi_\mathrm{L}}[0;\tau]
-
U_{\pi_\mathrm{H}}[0;\tau].
\end{aligned}
\]
The penultimate inequality follows from the definition of
$B^\star$ when the total variation is positive. When it is
zero, finiteness of $B^\star$ forces the cost differential
to be nonpositive, so the inequality still holds.

Thus $w$ satisfies incentive compatibility for every
$\tau\in\gT$. Its budget is at most $B^\star$, and the lower
bound above shows that it is budget-optimal with budget
exactly $B^\star$.

\section{Computing Contracts}

\subsection{Algorithm pseudocode}\label{alg:pseudocode}

Please consult Algorithm~\ref{alg:slack}.

\begin{algorithm}[H]
\caption{Computing a likelihood-ratio contract}
\label{alg:slack}
\renewcommand{\algorithmicrequire}{\textbf{Input:}}
\renewcommand{\algorithmicensure}{\textbf{Output:}}
\begin{algorithmic}[1]
\REQUIRE Duration $T$ and reachable-history representation;
threshold $\alpha$, accuracy $\varepsilon$; stopping-reward
oracle $R_B$, likelihood-ratio oracle $M$, and next-signal
probability oracle $\gP_{\pi_{\mathrm H}}$
\ENSURE Incentive-compatible contract within $\varepsilon$
of $B_\alpha^\star$, or a report that no finite budget exists
\STATE \textbf{Subroutine} \textsc{Verify}$(B)$
\STATE $V_B(\vs_{1:T})\gets R_B(\vs_{1:T})$
for all terminal histories
\FOR{$t=T-1,\ldots,1$}
    \FOR{each history $\vs_{1:t}$}
        \STATE $C\gets
        \sum_s\gP_{\pi_{\mathrm H}}(s\mid\vs_{1:t})
        V_B(\vs_{1:t},s)$
        \STATE $V_B(\vs_{1:t})\gets
        \max\{R_B(\vs_{1:t}),C\}$
    \ENDFOR
\ENDFOR
\RETURN
$V_B(\emptyset)\gets
\sum_{s_1}\gP_{\pi_{\mathrm H}}(s_1\mid\emptyset)V_B(s_1)$
\STATE \textbf{End subroutine}
\IF{$\neg$\textsc{CheckFiniteness}()}
    \RETURN ``no finite budget''
\ENDIF
\STATE Using \textsc{Verify}$(B)$ as a feasibility oracle,
find $U\in[B_\alpha^\star,B_\alpha^\star+\varepsilon]$
by doubling and binary search
\RETURN
$w_U(\vs_{1:t})
=
U\,\indicator{M(\vs_{1:t})\le1/\alpha}$
\end{algorithmic}
\end{algorithm}

\subsection{Proof of Theorem~\ref{thm:compute-budget}}
\label{app:algorithms}

Throughout, $\gT=\gT_T$ and the reachable signal tree
through $T$ is finite. For
$\pi\in\{\pi_{\mathrm H},\pi_{\mathrm L}\}$ and
$\tau\in\gT_T$, $\{\tau=t\}\in\sigma(\vs_{1:t})$.
Hence, by the tower property,
$\mathbb E_\pi[\widehat c_\pi(\vs_{1:\tau})]
=
\mathbb E_\pi[c(\vg_{1:\tau})]$.
Moreover,
$0\le c(\vg_{1:\tau})\le T\max_{g\in\gG}c_g$
and $0\le w_B(\vs_{1:\tau})\le B$, so all expected
costs and payments below are finite.

The stopped change-of-measure identity gives
$\mathbb E_{\pi_{\mathrm L}}
[\widehat c_{\pi_{\mathrm L}}(\vs_{1:\tau})]
=
\mathbb E_{\pi_{\mathrm H}}
[M(\vs_{1:\tau})
\widehat c_{\pi_{\mathrm L}}(\vs_{1:\tau})]$,
and analogously for the payment. Consequently,
\begin{equation}
\label{eq:appendix-reward-identity}
U_{\pi_{\mathrm L}}[w_B;\tau]
-
U_{\pi_{\mathrm H}}[w_B;\tau]
=
\mathbb E_{\pi_{\mathrm H}}[R_B(\vs_{1:\tau})].
\end{equation}

We next show that \textsc{Verify}$(B)$ returns the largest
incentive violation. At time $T$, stopping is mandatory, so
$V_B(\vs_{1:T})=R_B(\vs_{1:T})$. At a reachable history
$\vs_{1:t}$ with $t<T$, a rule either stops immediately or
continues. For any rule continuing beyond $t$, conditioning
on the next signal gives
\[
\mathbb E_{\pi_{\mathrm H}}
[R_B(\vs_{1:\tau})\mid\vs_{1:t}]
\le
\sum_s
\gP_{\pi_{\mathrm H}}(s\mid\vs_{1:t})
V_B(\vs_{1:t},s).
\]
Equality is attained by following, after observing $s$, an
optimal rule for the resulting history. Hence,
\begin{equation}
\label{eq:bellman-recursion}
V_B(\vs_{1:t})
=
\max\!\left\{
R_B(\vs_{1:t}),\
\sum_s
\gP_{\pi_{\mathrm H}}(s\mid\vs_{1:t})
V_B(\vs_{1:t},s)
\right\}.
\end{equation}
This is exactly the recursion in \textsc{Verify}$(B)$.
Backward induction also shows that a maximizing rule stops
at the first history where
$V_B(\vs_{1:t})=R_B(\vs_{1:t})$.
Averaging over the first signal gives
$V_B(\emptyset)
=
\max_{\tau\in\gT_T}
\mathbb E_{\pi_{\mathrm H}}[R_B(\vs_{1:\tau})]$.
Combining this with
Eq.~\ref{eq:appendix-reward-identity},
\[
V_B(\emptyset)
=
\max_{\tau\in\gT_T}
\bigl(
U_{\pi_{\mathrm L}}[w_B;\tau]
-
U_{\pi_{\mathrm H}}[w_B;\tau]
\bigr),
\]
so $w_B$ implements $\pi_{\mathrm H}$ exactly when
$V_B(\emptyset)\le0$.

Finally,
$\mathbb E_{\pi_{\mathrm H}}[M(\vs_{1:\tau})]=1$
implies
\[
\mathbb E_{\pi_{\mathrm H}}
[(1-M)\indicator{M\le1/\alpha}]
=
\mathbb E_{\pi_{\mathrm H}}
[(M-1)\indicator{M>1/\alpha}]
\ge0.
\]
Since
$w_B=B\indicator{M\le1/\alpha}$, the incentive violation
for each fixed $\tau$ is affine in $B$ with nonpositive
slope. Thus $V_B(\emptyset)$, as the maximum of finitely
many such affine functions, is convex, piecewise linear,
and nonincreasing.

Each backward pass visits every node and transition once,
so \textsc{Verify}$(B)$ requires $O(n)$ arithmetic
operations and oracle calls. Finiteness is handled below by
one additional backward pass. Once finiteness is established,
monotonicity justifies the doubling and binary search used
in Algorithm~\ref{alg:slack}.

\subsection{Finiteness check}
\label{app:lmax}

For $\tau\in\gT_T$, incentive compatibility is equivalent to
\[
B\,
\mathbb E_{\pi_{\mathrm H}}
\!\left[
(M(\vs_{1:\tau})-1)
\indicator{M(\vs_{1:\tau})>1/\alpha}
\right]
\ge
U_{\pi_{\mathrm L}}[0;\tau]
-
U_{\pi_{\mathrm H}}[0;\tau].
\]
The coefficient of $B$ is nonnegative and vanishes exactly
when
$\mathbb P_{\pi_{\mathrm H}}
(M(\vs_{1:\tau})>1/\alpha)=0$.
Hence a finite budget fails to exist precisely when there is
a stopping rule with
$M(\vs_{1:\tau})\le1/\alpha$ almost surely under
$\pi_{\mathrm H}$ while
$U_{\pi_{\mathrm L}}[0;\tau]
>
U_{\pi_{\mathrm H}}[0;\tau]$.
Since the tree admits only finitely many stopping rules,
\begin{equation}
\label{eq:finite-budget-criterion}
B_\alpha^\star<\infty
\quad\Longleftrightarrow\quad
\max_{\substack{\tau\in\gT_T\\
\mathbb P_{\pi_{\mathrm H}}
(M(\vs_{1:\tau})\le1/\alpha)=1}}
\bigl(
U_{\pi_{\mathrm L}}[0;\tau]
-
U_{\pi_{\mathrm H}}[0;\tau]
\bigr)
\le0,
\end{equation}
where the maximum of an empty set is $-\infty$.

The maximum in
Eq.~\ref{eq:finite-budget-criterion} is computed by the
same backward recursion as \textsc{Verify}, with $B=0$ and
modified stopping reward
\[
\widetilde R(\vs_{1:t})
=
\begin{cases}
R_0(\vs_{1:t}),
&
M(\vs_{1:t})\le1/\alpha,\\
-\infty,
&
M(\vs_{1:t})>1/\alpha.
\end{cases}
\]
Thus stopping is permitted only below the likelihood-ratio
threshold. The same backward-induction argument shows that
the value at the empty history equals the maximum in
Eq.~\ref{eq:finite-budget-criterion}. Accordingly,
\textsc{CheckFiniteness}() returns true exactly when this
value is nonpositive, requiring one additional $O(n)$
backward pass.

At $\alpha=1$, the coefficient of $B$ is the total-variation
distance between the stopped laws, so this reduces to the
zero-total-variation obstruction in
Theorem~\ref{thm:min-budget}.

\subsection{Numerically stable Bernoulli recursion}
The subroutine \textsc{Verify} evaluates the stopping reward
\[
R_B
=
\widehat c_{\pi_{\mathrm H}}
-
M\widehat c_{\pi_{\mathrm L}}
-
(1-M)B\phi
\]
at each state, where $\phi$ is a binary audit; for example,
the minimum-budget likelihood-ratio audit uses
$\phi(\vs_{1:t})=\indicator{M(\vs_{1:t})\le1}$.
In the Bernoulli setting, a state is represented by
$(t,\ell)$, where $\ell$ is the number of failures among the
first $t$ signals, and
\[
M_t(\ell)
=
\left(\frac{q}{p}\right)^{t-\ell}
\left(\frac{1-q}{1-p}\right)^\ell.
\]
Thus $M_t(\ell)$ can be exponentially large or small in $t$.
Directly forming the likelihood ratio can therefore overflow
or underflow in double precision well before the horizons of
interest. Rescaling the final output does not resolve this,
since the numerical error occurs inside the Bellman recursion.

We instead absorb the probability of reaching each state into
the value function. Let
\[
\nu_t(\ell)
=
\binom{t}{\ell}p^{t-\ell}(1-p)^\ell,
\qquad
\mu_t(\ell)
=
\binom{t}{\ell}q^{t-\ell}(1-q)^\ell
\]
denote the probabilities of state $(t,\ell)$ under
$\pi_{\mathrm H}$ and $\pi_{\mathrm L}$, respectively, and
define
\[
W_t(\ell):=\nu_t(\ell)V_t(\ell).
\]
The key identity is
$\nu_t(\ell)M_t(\ell)=\mu_t(\ell)$.
Hence, with constant per-round serving costs
$c_{\mathrm H}$ and $c_{\mathrm L}$, the scaled stopping
reward becomes
\[
\widehat R_B(t,\ell)
:=
\nu_t(\ell)R_B(t,\ell)
=
t\bigl(
c_{\mathrm H}\nu_t(\ell)
-
c_{\mathrm L}\mu_t(\ell)
\bigr)
-
B\bigl(
\nu_t(\ell)-\mu_t(\ell)
\bigr)\phi_t(\ell).
\]
This expression involves only probabilities bounded by one
and never requires forming $M_t(\ell)$ itself. For the
likelihood-ratio audit, even the test can be evaluated without
forming the ratio: at cutoff one,
$\phi_t(\ell)=\indicator{\mu_t(\ell)\le\nu_t(\ell)}$.

The Bellman recursion can be written directly in these scaled
variables. Since
\[
\frac{\nu_t(\ell)p}{\nu_{t+1}(\ell)}
=
\frac{t+1-\ell}{t+1},
\qquad
\frac{\nu_t(\ell)(1-p)}{\nu_{t+1}(\ell+1)}
=
\frac{\ell+1}{t+1},
\]
we obtain
\[
\nu_t(\ell)
\left[
pV_{t+1}(\ell)
+
(1-p)V_{t+1}(\ell+1)
\right]
=
\frac{
(t+1-\ell)W_{t+1}(\ell)
+
(\ell+1)W_{t+1}(\ell+1)
}{t+1}.
\]
Because $\nu_t(\ell)>0$, multiplying both the stopping and
continuation values by $\nu_t(\ell)$ does not change which is
larger. Thus the original Bellman recursion is equivalent to
\[
W_t(\ell)
=
\max\left\{
\widehat R_B(t,\ell),\
\frac{
(t+1-\ell)W_{t+1}(\ell)
+
(\ell+1)W_{t+1}(\ell+1)
}{t+1}
\right\},
\]
with terminal condition
$W_T(\ell)=\widehat R_B(T,\ell)$.
Finally,
\[
V_B(\emptyset)
=
pV_1(0)+(1-p)V_1(1)
=
W_1(0)+W_1(1).
\]
The scaled recursion therefore produces exactly the same
stopping decisions and root value as the original recursion,
while avoiding the exponentially large likelihood ratios
that cause numerical overflow.

\section{Betting Audits}
\label{app:betting}

\subsection{Service-level guarantee}\label{app:bet_service}

Suppose $s_t\in[a,b]$,
$0\le\lambda_t\le1/(b-s_0)$, and $\lambda_t$ is measurable
with respect to the past. Then
$1+\lambda_t(s_0-s_t)\ge0$. If
\[
\mathbb E_{\pi_{\mathrm H}}
[s_t\mid\mathcal F_{t-1}]
\ge s_0,
\]
then
\[
\begin{aligned}
\mathbb E_{\pi_{\mathrm H}}
\!\left[
1+\lambda_t(s_0-s_t)
\mid\mathcal F_{t-1}
\right]
&=
1+\lambda_t
\left(
s_0-
\mathbb E_{\pi_{\mathrm H}}
[s_t\mid\mathcal F_{t-1}]
\right) \le1.
\end{aligned}
\]
Thus
\[
\mathbb E_{\pi_{\mathrm H}}
[E_t\mid\mathcal F_{t-1}]
\le E_{t-1},
\]
so $(E_t)$ is a nonnegative supermartingale with $E_0=1$.

Ville's inequality gives
\[
\mathbb P_{\pi_{\mathrm H}}
\left(
\sup_tE_t\ge\frac1\alpha
\right)
\le\alpha.
\]
Since
$\{E_\tau>1/\alpha\}
\subseteq
\{\sup_tE_t\ge1/\alpha\}$
for every stopping time $\tau$,
\[
\mathbb P_{\pi_{\mathrm H}}
(E_\tau>1/\alpha)
\le\alpha,
\]
or equivalently
\[
\mathbb P_{\pi_{\mathrm H}}
(E_\tau\le1/\alpha)
\ge1-\alpha.
\]

\subsection{Likelihood-ratio interpretation}\label{app:bet_lr}

Assume now that
\[
\mathbb E_{\pi_{\mathrm H}}
[s_t\mid\mathcal F_{t-1}]
=
s_0.
\]
For any admissible predictable $\lambda_t$, define, at a
reachable history $h$,
\[
\gP_t^\lambda(ds\mid h)
=
\bigl(1+\lambda_t(h)(s_0-s)\bigr)
\gP_{\pi_{\mathrm H}}(ds\mid h).
\]
Nonnegativity follows from the restriction on $\lambda_t$,
while
\[
\int
\bigl(1+\lambda_t(h)(s_0-s)\bigr)
\gP_{\pi_{\mathrm H}}(ds\mid h)
=
1
\]
by the conditional-mean assumption. Hence
$\gP_t^\lambda(\cdot\mid h)$ is a probability kernel and
\[
\frac{
d\gP_t^\lambda(\cdot\mid h)
}{
d\gP_{\pi_{\mathrm H}}(\cdot\mid h)
}(s)
=
1+\lambda_t(h)(s_0-s).
\]

The chain rule for likelihood ratios therefore gives
\[
\frac{
d\gP_{1:t}^\lambda
}{
d\gP_{\pi_{\mathrm H},1:t}
}(\vs_{1:t})
=
\prod_{i=1}^t
\bigl(1+\lambda_i(s_0-s_i)\bigr)
=
E_t.
\]
Thus $E_t$ is exactly the likelihood-ratio process between the
promised signal law and the alternative law induced by the
chosen stakes.

For Bernoulli signals, let the promised and alternative laws
be $\operatorname{Ber}(p)$ and $\operatorname{Ber}(q)$,
respectively, with $q<p$. Taking $s_0=p$ and
\[
\lambda
=
\frac{p-q}{p(1-p)}
\]
yields
\[
1+\lambda(p-1)=\frac qp,
\qquad
1+\lambda p=\frac{1-q}{1-p},
\]
which are the one-step likelihood ratios for success and
failure. Hence the corresponding betting process equals the
Bernoulli likelihood-ratio process exactly.

\subsection{Finding the minimal IC budget for a betting contract}\label{app:bet_pf}

Let $\mathcal B$ denote the set of budgets $B\ge0$ for which $w^{\mathrm{bet}}_{\alpha,B}$ is incentive compatible, assume $\mathcal B\neq\emptyset$, and let $\textsc{Verify}_\phi$ denote the backward pass of Section 4.1 with $\indicator {M\le1/\alpha}$ replaced by $\phi$.

The reward identity~\eqref{eq:appendix-reward-identity} and the backward-induction argument in the proof of Theorem~\ref{thm:compute-budget} use only that the payment is a function of the observed history, so they apply verbatim to $w^{\mathrm{bet}}_{\alpha,B}=B\phi$. Hence $\textsc{Verify}_\phi(B)=\max_{\tau\in\gT_T}\bigl(U_{\pi_{\mathrm L}}[w^{\mathrm{bet}}_{\alpha,B};\tau]-U_{\pi_{\mathrm H}}[w^{\mathrm{bet}}_{\alpha,B};\tau]\bigr)$, and $B\in\gB$ if and only if $\textsc{Verify}_\phi(B)\le0$. The audit is evaluated along the tree via $E(h,s)=E(h)\bigl(1+\lambda(h)(s_0-s)\bigr)$, so each pass takes $O(n)$ operations.

By the stopped change of measure, for each $\tau\in\gT_T$,
\begin{equation}
\label{eq:bet-affine}
\begin{aligned}
U_{\pi_{\mathrm L}}[w^{\mathrm{bet}}_{\alpha,B};\tau]&-U_{\pi_{\mathrm H}}[w^{\mathrm{bet}}_{\alpha,B};\tau]
=U_{\pi_{\mathrm L}}[0;\tau]-U_{\pi_{\mathrm H}}[0;\tau]\\
&-B\,\mathbb E_{\pi_{\mathrm H}}\bigl[\bigl(1-M(\vs_{1:\tau})\bigr)\,\phi(\vs_{1:\tau})\bigr],
\end{aligned}
\end{equation}
which is affine in $B$. Since $\gT_T$ is finite, $\textsc{Verify}_\phi$ is a maximum of finitely many affine functions of $B$, hence convex and continuous. Thus $\gB$ is a closed interval, and, being nonempty and bounded below by $0$, it has a least element $B^{\mathrm{bet}}$.

Therefore, given any element $\tilde{B}$ of this interval, bisection finds the minimum one with $\varepsilon$ precision in $O(\log(\tilde{B} / \varepsilon))$ tries.

Suppose that deviating saves costs under every admissible stopping rule, that is, $U_{\pi_{\mathrm L}}[0;\tau]>U_{\pi_{\mathrm H}}[0;\tau]$ for all $\tau\in\gT_T$. By Eq.~\ref{eq:bet-affine}, if $\mathbb E_{\pi_{\mathrm H}}[(1-M(\vs_{1:\tau}))\phi(\vs_{1:\tau})]\le0$ for some $\tau$, then $w^{\mathrm{bet}}_{\alpha,B}$ violates incentive compatibility at $\tau$ for every $B\ge0$, while for every $\tau$ with a positive expectation the constraint at $\tau$ is a lower bound on $B$. Consequently, under the assumption of $U_{\pi_{\mathrm L}}[0;\tau]>U_{\pi_{\mathrm H}}[0;\tau]$, we have that $\gB\neq\emptyset$ if and only if
\[
\min_{\tau\in\gT_T}\mathbb E_{\pi_{\mathrm H}}\bigl[\bigl(1-M(\vs_{1:\tau})\bigr)\,\phi(\vs_{1:\tau})\bigr]>0,
\]
and in that case $\gB=[B^{\mathrm{bet}},\infty)$, so doubling from $\varepsilon$ reaches a feasible budget after $O(\log(B^{\mathrm{bet}}/\varepsilon))$ steps, after which the bisection above applies. Both checks take one backward pass each. The minimum above is computed by the recursion~\eqref{eq:bellman-recursion} with stopping reward $-(1-M)\phi$. For the cost condition, the reward identity with zero payment gives $U_{\pi_{\mathrm L}}[0;\tau]-U_{\pi_{\mathrm H}}[0;\tau]=\mathbb E_{\pi_{\mathrm H}}[R_0(\vs_{1:\tau})]$, where $R_0=\widehat c_{\pi_{\mathrm H}}-M\,\widehat c_{\pi_{\mathrm L}}$ is the stopping reward of Section~\ref{sec:algorithms} at $B=0$, so its minimum over $\tau$ is computed by the same recursion with stopping reward $-R_0$.

We note that computing $B^{\mathrm{bet}}$, as well as verifying the conditions above, requires the signal laws induced by both policies and their costs, through $\gP_{\pi_{\mathrm H}}$, $M$ and $\widehat c_\pi$. Only the audit itself is distribution-free: $E_t$ is evaluated from the observed signals alone, and its level-$\alpha$ guarantee relies only on the service-level condition.

\section{Empirical Evaluation}\label{app:reproduction}

All computations are run on a personal computer.

\subsection{Qualitative Analysis}

\paragraph{Reproducibility constants for min-budget and min-pay}

For Figure~\ref{fig:synthetic} (Left), we use
$p=\tfrac12$, $q=\tfrac14$, $c_{\mathrm H}=1$,
$c_{\mathrm L}=0$, and $T=3$. The min-budget contract is
computed using Theorem~\ref{thm:min-budget}, while the min-pay
contract is obtained from the linear program in
Appendix~\ref{app:lp}.

\paragraph{Reproducibility constants for the deterred set}

For Figure~\ref{fig:synthetic} (Center), the promised router is
$(p,c_{\mathrm H})=(0.9,1)$, the reference deviation is
$(q,c_{\mathrm L})=(0.4,0.5)$, and $T=3$. The displayed region
is obtained by evaluating the resulting contract against each
candidate deviation $(q',c')$.

\paragraph{Reproducibility constants for the worst-case stopping rule}

For Figure~\ref{fig:synthetic} (Right), we use
$p=0.8$, $q=0.2$, $c_{\mathrm H}=1$, $c_{\mathrm L}=0$,
and $T=10$. The stopping rule shown is the maximizing rule
returned by the dynamic program of Section~\ref{sec:algorithms}.

\subsection{Practical Considerations}

We take accuracies and costs from the SWE-bench Verified Mini board of the
Holistic Agent Leaderboard \citep[HAL;][]{kapoor2026hal,hal2026swebenchmini}, read directly from its published table and retrieved on 4 October 2026. The benchmark is a $50$-task subset of SWE-bench Verified \citep{jimenez2024swebench}, and we identify one task with one step of the subscription. We keep the SWE-Agent entries throughout, so that two routers differ only in the model they serve. An entry's reported accuracy is the fraction of the $50$ tasks it resolves, which we read as the per-step probability of a passing evaluation; $p$ and
$q$ are therefore the accuracies of the promised and the deviating router.
Its reported cost is the total API cost of running the agent over all $50$
tasks, so we set $c_{\mathrm H}$ and $c_{\mathrm L}$ to those totals divided by $50$, an estimate of the expected cost of a single step.

\paragraph{Reproducibility constants for the role of $\alpha$}

For Figure~\ref{fig:hal} (Left), the promised model is
Claude Opus 4.1 and the reference deviation is Claude 3.7
Sonnet, giving
$(p,q,c_{\mathrm H},c_{\mathrm L})
=(0.61,0.50,27.03,8.05)$.
We use horizons $T\in\{25,50,100,200\}$ and vary $\alpha$
over the range shown in the figure.

\paragraph{Reproducibility constants for partial information}

For Figure~\ref{fig:hal} (Center), we use the same model pair
with $T=25$ and reference service level $s_0=p=0.61$.
We vary the betting stake $\lambda$; the stake reproducing
the likelihood-ratio audit is
$\lambda^\star=(1-q/p)/(1-p)$.

\paragraph{Reproducibility constants for deterring a family of models}

For Figure~\ref{fig:hal} (Right), the promised model is
Claude Sonnet 4.5 High and the reference deviation is
Claude Sonnet 3.7 High, with
$(p,q,c_{\mathrm H},c_{\mathrm L})
=(0.72,0.54,9.28,7.78)$
and $T=25$. The resulting contract is evaluated against each
of the remaining Claude routers in the leaderboard.

\end{document}